# Seeding ice growth at ambient conditions using nano graphene oxide

Yi Zheng[1], Chengliang Su[1], Jiong Lu[1] and Kian Ping Loh[1*]

[1]Department of Chemistry and Graphene Research Centre, 3 Science Drive 3, National University of Singapore, Singapore 117543

[*]Correspondence to: chmlohkp@nus.edu.sg

**Water wetting on a hydrophobic surface at ambient conditions is disallowed by the non-polar nature of the surface and high vapor pressure of water (*1–3*). However, the presence of sub-millimeter sized hydrophilic patches allows the waxy wings of desert beetles to become wettable by morning mist (*4, 5*). Here, we show that a sprinkle of graphene oxide nanoflakes (nanoGOs) is effective in condensing water nanodroplets and seeding ice epitaxy on graphite at ambient conditions. By controlling relative humidity and nanoGO density, we are able to study the formation of a complete ice wetting layer on a time scale of 20 hours. This presents an unprecedented opportunity to visualize ice nucleation and growth in real time using non-contact atomic force microscopy. The stages of crystallization, as proposed by Ostwald in 1897 (*6*), is fully unfolded at a microscopic level for the first time. We obtain real-time imaging of sequential phase transition from amorphous ice to a transient cubic ice $I_c$ stage and finally to the stable hexagonal ice $I_h$. Most interestingly, we discover that ice nucleation and growth can be influenced by modifying the functional groups of nanoGO, and by intermolecular hydrogen-bonding between nanoGOs. This affords a strategy to control heterogenous ice nucleation and snow crystal formation (*7, 8*).**

The interaction of water with solid surfaces is one of the most pervasive natural phenomena which underpin rain precipitation, snow formation, rock erosion *etc*. The wetting of surfaces by ambient water is also of crucial importance in many processes like heterogeneous catalysis, photocatalysis, microelectronics and drug development (*1,3*). In bulk ice which is formed below 0 ˚C, water molecules pack in a hexagonal ice $I_h$ structure with four hydrogen bonds arranged in tetrahedral geometry (*8*). Above 0 ˚C, the randomization of hydrogen bonds through concerted bond breaking and formation in the order of picoseconds leads to the liquid phase (*9,10*). A new paradigm emerges recently that the phase transition of liquid water and ice at the freezing point may not apply in the two dimensional (2D) limit. Recent research revealed that in the presence of a hydrophilic interface (*11*), or by nano-confinement (*12,13*), ice $I_h$ wetting layers can persist even at ambient conditions. To date, insights into the hydration structure and wetting dynamics of ambient water on solids is derived mainly from studies on single-crystal metal surfaces carried out at cryogenic temperatures in vacuum conditions (*1,3,14*). Molecular level studies of water/ice nucleation and growth on solid surface at ambient conditions remain formidable due to the highly mobile nature of water molecules and its high vapor pressure (*2*).

In our ice growth experiments, we use graphene oxide nanoflakes (nanoGOs) as ice nucleation seeds. These nanoGOs were base-refluxed and were characterized by a higher degree of restored $sp^2$ conjugation compared to as-synthesized nanoGOs and the presence of carboxylate ($COO^-$) groups on its periphery, as evidenced by vibrational spectroscopy (Supplementary Fig. 1). After spin coating on freshly peeled Highly Ordered Pyrolytic Graphite (HOPG) surface, the samples are dried at 80 °C for 15 minutes before non-contact atomic force microscopy (NCAFM) imaging. The typical NCAFM oscillating amplitude for direct imaging of ice epitaxial domains is 2 nm. Using higher oscillating amplitude results in a perturbation of the wetting layer (see Supplementary Fig. 2 for an amplitude setpoint of 6 nm) or even penetration of the water films (>10 nm; Supplementary Fig. 2). The latter allows us to directly image the underlying substrate and determine the physical dimensions of nanoGOs, which have heights varying from 0.5 nm to several nm (Supplementary Fig. 2).

Figure 1a shows a *Stenocara* beetle in Namib desert bathing in the morning mist. The hydrophobic wings of the beetle become wettable in the presence of sub-millimeter hydrophilic protrusions. Taking the cue from nature, we moderate graphite substrate with 1-10% of hydrophilic nanoGOs. The graphite surface is atomically flat which does not disrupt the fragile hydrogen bonding in ice structures, and it can quickly conduct the latent heat produced by ice condensation. The triangular sublattice of graphite (2.46 Å) matches the natural ice structure very well. All these factors favor the epitaxial growth of a commensurate ($\sqrt{3}\times\sqrt{3}$)R30° ice $I_h$ overlayer (*15,16*). However, at ambient conditions, the formation of 2D ice overlayers on graphite is suppressed by strong sublimation due to the high vapor pressure of water (19.5 Torr)

(*2*). This difficulty can be circumvented by fast surface diffusion of water molecules (*17*) from scattered nanodroplet reservoirs, which are readily clustered around nanoGOs at relative humidity (RH) as low as 15%. As illustrated in Fig. 1b, right after the baking procedure, liquid droplets can be efficiently captured by the strongly hydrophilic carboxylate groups residing in nanoGOs. Surface tension leads to the formation of near-spherical nanodroplets covering the underlying nanoGOs, with a typical liquid thickness of ~0.5 nm (Supplementary Fig. 2). The negatively charged carboxylate groups provide the first anchor site for ice nucleation (Fig. 1b). From these, the growth of ice crystals is driven by the outward diffusion of water from the nanoGOs. The final coverage of ice domains is a dynamical balance between water adsorption on nanoGOs and sublimation on graphite surface (Fig. 1b), which can be controlled by tuning RH and nanoflake density. At high humidity levels (RH>70%), the graphite surface is quickly covered by ice crystalline domains in a few minutes. Fig. 1c shows a characteristic NCAFM image of large-area ice thin film formed on HOPG. It has to be emphasized that here, the surface of ice overlayers is imaged directly, in contrast to the recent report of nanoconfined ice sandwiched between graphene and mica (*13*). The direct imaging provides molecular-level lateral resolution of ice overlayers and allows the nucleation and growth process to be dynamically studied. For instance, we are able to observe reversible phase transition between an unconventional 1D ice structure and ice $I_h$ (left panel of Fig. 1d). At a high nanoGO coverage on the graphite surface, we can observe a non-reversible staged crystallization in multilayer ice growth (right panel of Fig. 1d), which is fully unfolded for the first time since proposed by Ostwald in 1897 (*6*).

Fig. 2a shows a typical NCAFM image of the first ice wetting layer formed on HOPG, seeded by low density nanoGOs (~0.9% coverage) at a low RH of 16%. These ice domains are ca. 0.3 nm in heights and consist of periodical 1D chains (the inset of Fig. 2a). The measured step height is significantly smaller than the optimized distance between an ice $I_h$ monolayer and graphite (0.38 nm) (*15*), but agrees with an ice overlayer with water molecules lying flat on the surface. The formation of a flat-lying phase at low RH can be rationalized by the maximization of H-bonding between neighboring water molecules to stabilize ice domains against increasing sublimation. Unlike the tetrahedral H-bonding in ice $I_h$ which can extend infinitely (3D icing; see Supplementary Fig. 3), such coplanar adsorption breaks translational symmetry and can only form clusters or narrow stripes, determined by the "2D ice rule" (*18*). As illustrated in Fig. 1d, beyond the first flat-lying hexamer, adding each hexagon will introduce one defective double-acceptor molecules. Energy minimization gives rise to the periodical 1D chain structure with defect molecules on the periphery. The orientations of the 1D chains are characterized by a pronounced 3-fold symmetry (dashed lines in Fig. 2a). The symmetry originates in the coupling between nanoGOs and graphite, which becomes less likely as the size of nanoGOs increases. Indeed, we found that ice growth is rarely initiated by GO flakes larger than 100 nm (Supplementary Fig. 4). By increasing RH to 30%, isolated ice domains surrounding nanoGOs expanded and formed a continuous wetting layer (Fig. 2b). Unlike the case in Fig. 2a, the wetting layer at 30% RH has predominantly smooth surfaces although some local areas retain the periodical 1D chain structure. With a further increase in RH to 55%, graphite is fully covered by the wetting layer except for some non-coalescent pinholes (Fig. 2c). The step height of the smooth domains is consistent with one ice $I_h$ monolayer on HOPG (0.38 ± 0.02 nm; Fig. 2d). Phase image also confirms that this new phase is the first ice $I_h$ monolayer, as manifested by its distinct phase contrast from graphite (Supplementary Fig. 5). The observation implies that at higher humidity, the ice $I_h$ structure becomes kinetically favored due to higher packing density of water molecules. Above 55% RH, a second coplanar adlayer can be observed occasionally. This second layer is hydrogen bonded to the underlying ice wetting layer as indicated by the measured step height of 0.38 nm (Fig. 2e and 2f). However, in general, we did not observe the full coalescence of 1D chains into ice $I_h$ monolayer.

Although the phase transition between 1D ice and ice $I_h$ monolayer is driven kinetically by packing density and is reversible, the natural ice structure becomes energetically favorable once multilayer ice are formed. This is attributed to the formation of interlayer H-bonding, which compensates the energy cost of vertical buckling of ~0.1 nm between neighboring water molecules (*3*). Multilayer ice can be grown on graphite by increasing nanoGO coverage to enhance the water adsorption flux. At an intermediate RH of 40% and at nanoGO density of 4%, it takes 24 hours to form a continuous ice thin film, which allows the real-time wetting dynamics to be recorded by NCAFM (20 mins per image). As shown in Fig. 3a-e, the growth of ice multilayers on graphite under these conditions is highly anisotropic. The fast growth directions of each ice crystal follow the 3-fold symmetry as determined by the underlying graphite domains (dashed lines in Fig. 3a).

By tracking the step heights of individual ice crystals as a function of time, we found that in the beginning, the majority of freshly nucleated crystals are amorphous solid water (ASW), a different polymorph from ice $I_h$. This is manifested by a 0.68 ± 0.02 nm step height which disagrees with multiples of ice $I_h$ monolayer (Fig. 3f and raw data in Supplementary Fig. 6). This polymorph contains voids and dangling hydrogen bonds (Fig. 1d), which has minimum activation energy for nucleation due to structural similarity to the liquid phase (*14*). The ASW phase slowly develops within 1.5 hours into another metastable phase with a step height of 0.56 ± 0.02 nm (Fig. 3f and Supplementary Fig. 6). Noticeably, the thickness of 0.56 nm corresponds to two hexagonal monolayers of ice $I_c$ phase (*19,20*). Consistent with the water freezing experiments (*19*), such ice $I_c$ phase is transient and quickly evolves into ice $I_h$ within half an hour (Fig. 3f and Supplementary Fig. 6). The staged crystallization from the less stable polymorphs (ASW followed by ice $I_c$) to the most stable ice $I_h$ has been proposed by Ostwald as an empirical rule in 1897 (*6*), and is related to the nucleation kinetics of solid phase from liquid phase (*21, 22*).

The presence of carboxylate groups on the edges of nanoGO plays a key role in ice nucleation. This may be due to the enhanced electrostatic interactions between the charged H-acceptor $COO^-$ and water molecules, a phenomenon known as "charge-assisted H-bonding" (*23*). Indeed, we are able to suppress the nucleation ability of nanoGOs by charge neutralization, as demonstrated in Fig. 4. In $COO^-$, the negative charge is delocalized in between two oxygen atoms in a resonance structure, creating two strong H-bond acceptors (Fig. 4a). The resulting ice nucleation starts immediately after the baking process, and the following ice growth is highly anisotropic (Fig. 4a). In contrast, charge neutral carboxylic groups (by acid treatment) and ammonium carboxylates (by ammonia solution) are much weaker in initiating ice crystal growth, even when the density of nanoGO coverages are significantly increased (Fig. 4b and Fig. 4c). By allowing nanoGOs to self-assemble on graphite *via* hydrogen bonding, the anisotropic ice nucleation can be transformed into quasi-isotropic. As shown in Fig. 4d, nanoGOs can form 1D wires and seed ice growth surrounding the wire backbones. In contrast, the self-assembled ring structures do not favor ice nucleation. This can be explained by the fact that 1D H-bonding of carboxylic groups provides a building block with structural similarity to ice hexamer, which effectively promotes ice nucleation (see the schematic in Fig. 4d). However, the intermolecular association of ring structures in general does not provide such active sites for ice nucleation. For example, the strong linear H-bonding in a triple-carboxylic ring makes the C=O acceptor disfavors a second H-bonding, which requires a change in the acceptor directionality from linear to about 120˚ (see another example of a six-carboxylic ring in Supplementary Fig. 7).

It is remarkable that very low surface coverage of GO nanoflakes can transform the water wetting dynamics on a macroscopically large hydrophobic surface and effectively promote ice formation even at ambient conditions. This result has obvious implications for engineering water wetting on a wide range of surfaces for anti-fogging or anti-fouling applications (*24*). At a fundamental level, we show that the hybrid nanoGO-graphite template can be used as a model system to study crystal nucleation and growth in real time. By engineering the functional groups on nanoGOs and introducing different chemical modification, the long-debated heterogeneous ice nucleation and snow crystal formation could be better understood on a molecular level. Finally, the implication for graphene (*25, 26*) electronics is that the formation of ice-like clusters may occur near defects and oxygenated adsorbates in humid conditions. These may have critical influence on electronic properties.

**Acknowledgements:** Loh KP acknowledges funding support from Singapore Millenium Foundation Research Horizon Award R-143-000-417-133 as well as Economic Development Board (SPORE, COY-15-EWI-RCFSA/N197-1). Discussion with Prof A.H. Castro Neto is gratefully acknowledged.

# Figures

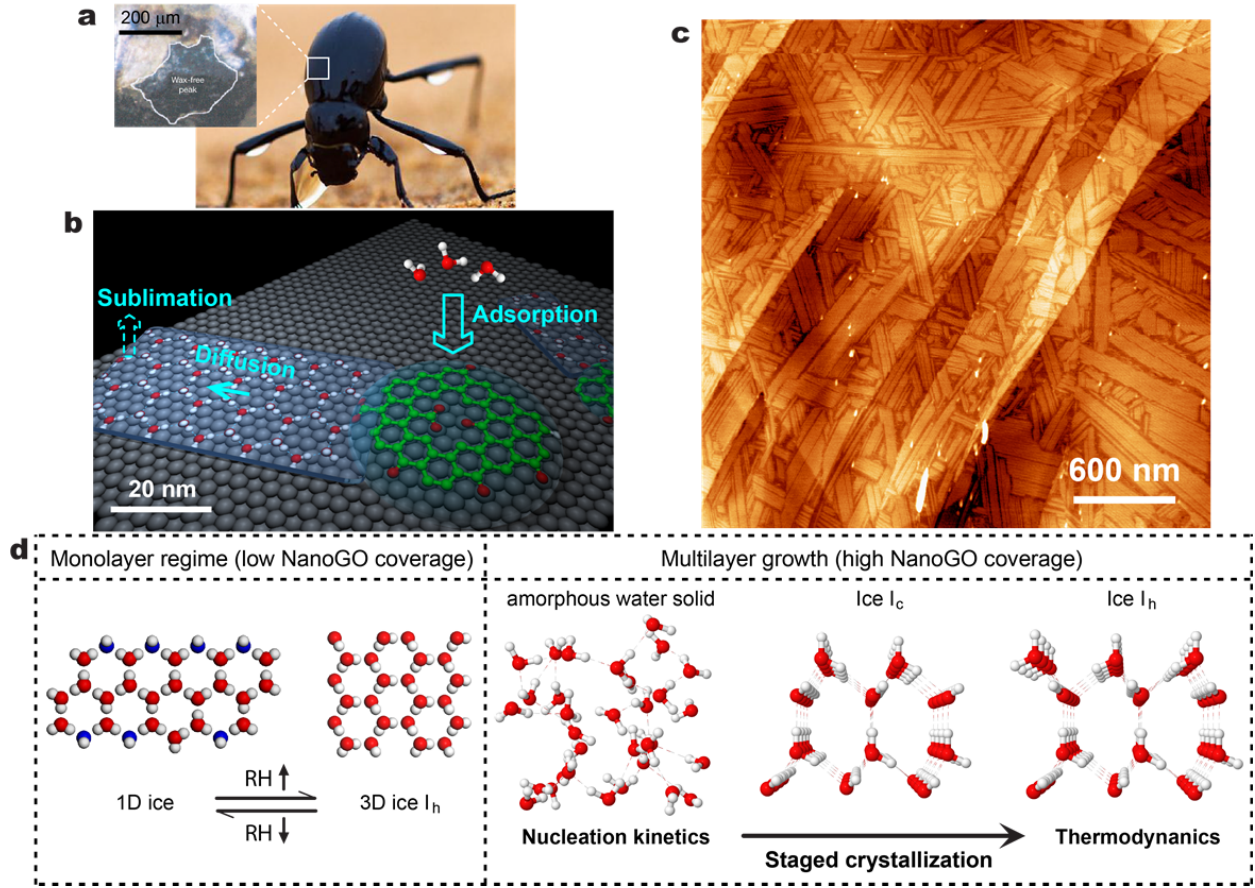

Figure 1: Ice formation on a hydrophobic surface at ambient conditions seeded by superhydrophilic nanoGOs. **a**, A *Stenocara* beetle in Namib desert baths in the morning mist. Inset: hydrophilic bumps embedded in the hydrophobic wings. Image reproduced with permission from ref. 4, ©2001 Nature. **b**, Schematic of ice formation on graphite, nucleated by carboxylate functional groups on nanoGOs. The coverage of ice domains on graphite is determined by dynamical balance between adsorption on nanoGOs and sublimation on graphite. **c**, High coverage of ice wetting layers formed on HOPG in few minutes due to exposure to high humidity levels. **d**, Phase diagram of ice formation on graphite as a function of RH and nanoGO density.

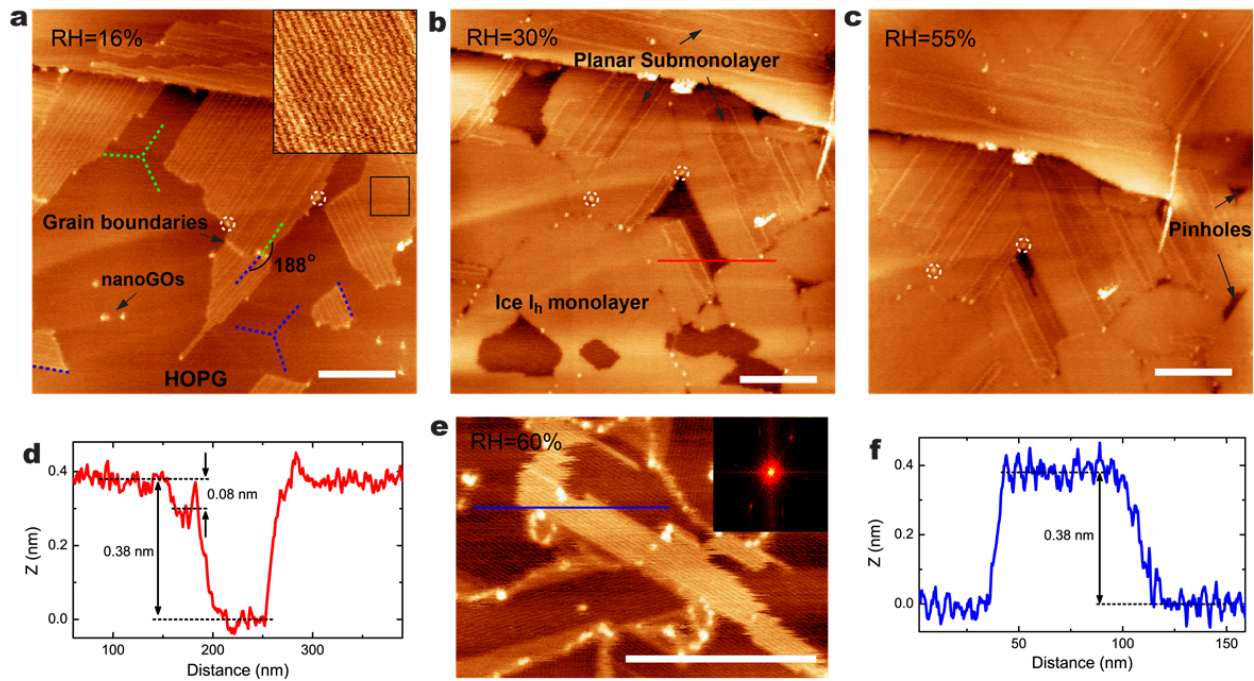

Figure 2: The first ice wetting layer on HOPG at ambient conditions seeded by superhydrophilic nanoGOs. **a**, Submonolayer ice wetting domains at 16% RH. Inset: zoom-in of the periodical 1D chain structure. Note that the coalescence of two misoriented ice domains leads to the formation of grain boundaries as indicated by the black arrow. Green and blue colors highlight two crystallographic directions which are misoriented by 8°. **b**, Faceted ice $I_h$ domains formation on HOPG with larger adsorption flux at 30% RH. **c**, At 55% RH, HOPG is nearly fully covered by the wetting layer. **d**, Height profile across the first ice wetting layer and HOPG, showing 0.08 nm difference between ice $I_h$ and planar submonolayer domains. **e**, Formation of a second coplanar submonolayer, hydrogen bonded to the underlying ice domain. **f**, Height profile across the 1st and 2nd planar submonolayers. Scale bars correspond to 200 nm. Dashed white circles highlight two nanoGOs as the coordination reference points.

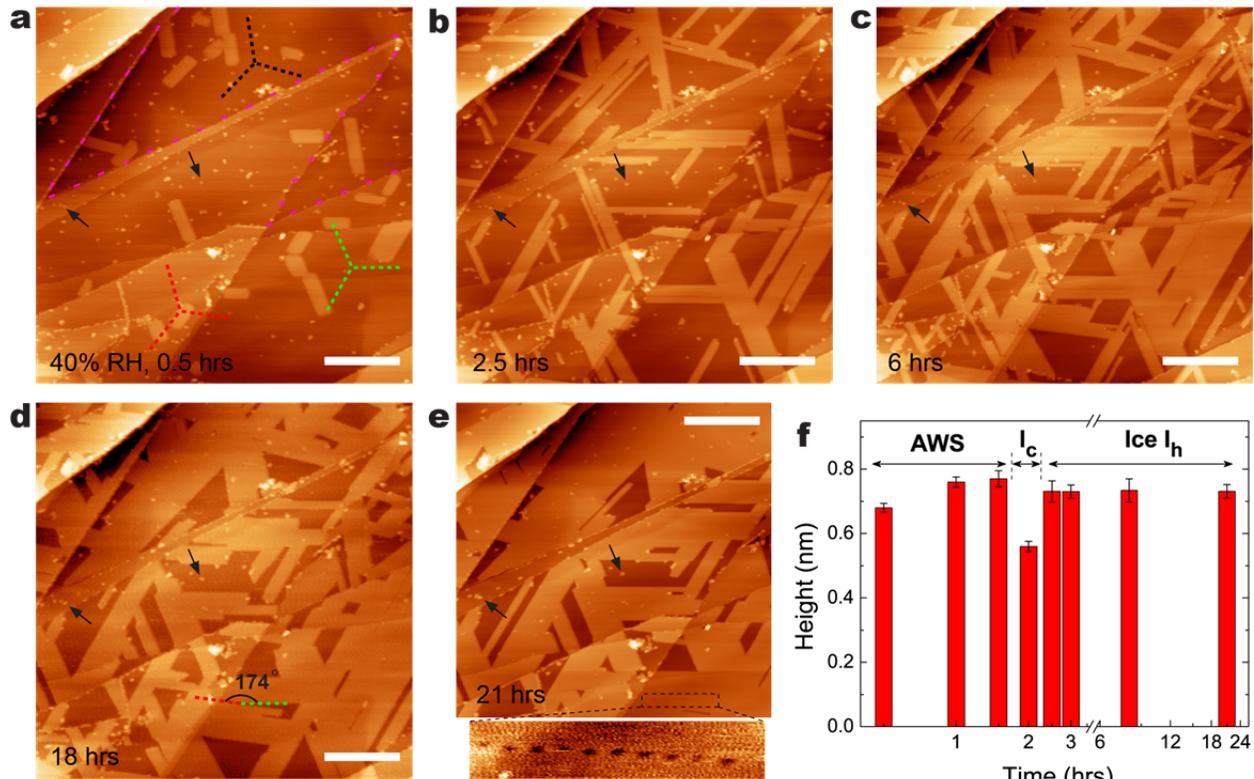

Figure 3: Real-time ice wetting dynamics on HOPG at 40% RH, seeded by 4% of nanoGOs. **a**, NCAFM image right after the baking, showing 3-fold ice crystal growth in registry to the underlying graphite lattices. Three crystallographic directions are determined using the orientation of ice crystals. **b**, NCAFM image after 2.5 hours. **c**, 6 hours. **d**, 18 hrs. **e**, 21 hours. Note that the coalescence of ice domains on the same graphite lattice creates triangular shaped pinholes. Bottom: Molecular-resolution imaging reveals the coalescence of two small mismatched ice domains by forming one dimensional pinhole arrays. **f**, Step height analysis of an ice crystalline domain as a function of time, showing polymorphism evolution from amorphous solid to ice $I_c$, and finally to ice $I_h$. Scale bars correspond to 500 nm.

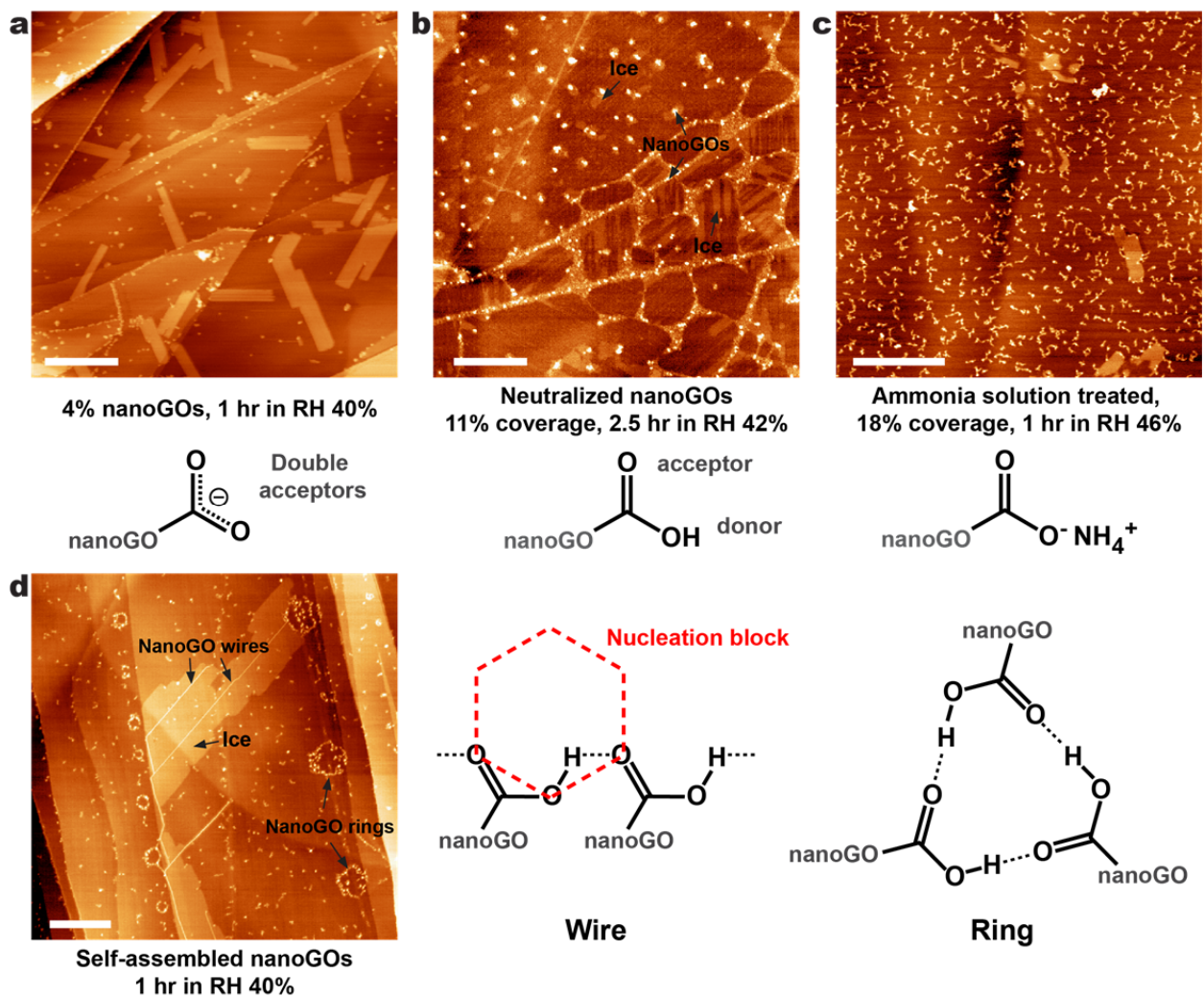

Figure 4: Tuning ice nucleation and growth behavior by modifying the acidity of the functional groups and by self-assembly of nanoGOs. **a**, Typical anisotropic ice growth nucleated by negatively charged carboxylates. **b**, After acid neutralization, the ice nucleation is suppressed. **c**, Charge neutralization by ammonia solution treatment of nanoGOs. The result is consistent with the acid treatment. **d**, Quasi-isotropic ice growth nucleated by self-assembled 1D nanoGO wires. Schematics on the right side show the possible H-bonding configuration in 1D wires and carboxylic rings. Scale bars correspond to 500 nm.

**Supplementary Materials:**

Materials and Methods

Supplementary Figures 1-8

References (*1-2*)